\documentclass[10pt,conference]{IEEEtran}

\pdfoutput=1

\usepackage{xspace}
\usepackage{paralist}
\usepackage{graphicx}
\usepackage{enumerate}
\usepackage{booktabs}
\usepackage{pdfpages}
\usepackage{mdframed}
\usepackage{subfigure}
\usepackage{flushend}
\usepackage{listings}
\usepackage{courier}
\usepackage{multirow}
\usepackage{tikz}
\usepackage{afterpage}
\usepackage{tabularx}

\usepackage[pdftex,colorlinks=true,pdfstartview=FitV,linkcolor=black,citecolor=black,urlcolor=black,bookmarks=false]{hyperref}
\usepackage{needspace}

\usepackage{ifthen}
\usepackage[normalem]{ulem} 
\usepackage{xcolor}

\newboolean{showedits}
\setboolean{showedits}{false} 
\setboolean{showedits}{true} 
\ifthenelse{\boolean{showedits}}
{
	\newcommand{\del}[1]{\textcolor{red}{\sout{#1}}} 
	\newcommand{\nbe}[3]{
		{\colorbox{#3}{\bfseries\sffamily\scriptsize\textcolor{white}{#1}}}
		{\textcolor{#3}{\sf\small$\blacktriangleright$\textit{#2}$\blacktriangleleft$}}}
}{
	\newcommand{\del}[1]{} 
	
	\newcommand{\nbe}[3]{}
}
\definecolor{mircolor}{rgb}{0.4,0.6,0.2}

\usepackage[most]{tcolorbox}
\ifthenelse{\boolean{showedits}}
{
  \newtcolorbox{inserted}{%
       title=Inserted text:,
       colframe=blue,colback=blue!5!white,
       breakable,
       leftrule=0mm, 
       bottomrule=0mm,
       rightrule=0mm,
       toprule=0mm,
       arc=0mm, outer arc=0mm,
       oversize
  }
  \newtcolorbox{deleted}{%
       title=Deleted text:,
       colframe=red,colback=red!5!white,
       breakable,
       leftrule=0mm, 
       bottomrule=0mm,
       rightrule=0mm,
       toprule=0mm,
       arc=0mm, outer arc=0mm,
       oversize
  }
  \newtcolorbox{refactored}{%
       title=Rewritten text:,
       colframe=blue,colback=red!5!white,
       breakable,
       leftrule=0mm, 
       bottomrule=0mm,
       rightrule=0mm,
       toprule=0mm,
       arc=0mm, outer arc=0mm,
       oversize
  }
}{

}
\usepackage{amssymb}
\newboolean{showcomments}
\setboolean{showcomments}{false}
\setboolean{showcomments}{true}

\ifthenelse{\boolean{showcomments}}
{\newcommand{\nbc}[3]{
 {\colorbox{#3}{\bfseries\sffamily\scriptsize\textcolor{white}{#1}}}
 {\textcolor{#3}{\sf\small$\blacktriangleright$\textit{#2}$\blacktriangleleft$}}}
 }
{\newcommand{\nbc}[3]{}
 }

\newcommand{\ie}{\emph{i.e.},\xspace}
\newcommand{\eg}{\emph{e.g.},\xspace}
\newcommand{\etal}{\emph{et al.}\xspace}
\newcommand{\etc}{\emph{etc.}\xspace}

\begin{document}

\title{The Impact of Feature Selection on Predicting the Number of Bugs}

\author{
\IEEEauthorblockN{Haidar Osman\textsuperscript{1}, Mohammad Ghafari\textsuperscript{2}, Oscar Nierstrasz\textsuperscript{2}}
\IEEEauthorblockA{
\textsuperscript{1}Swisscom AG, Switzerland\\
\textsuperscript{2}Software Composition Group, University of Bern, Switzerland\\
\{osman, ghafari, oscar\}@inf.unibe.ch}
}

\maketitle

\begin{abstract}
Bug prediction is the process of training a machine learning model on software metrics and fault information to predict bugs in software entities.
While feature selection is an important step in building a robust prediction model, there is insufficient evidence about its impact on predicting the number of bugs in software systems.
We study the impact of both correlation-based feature selection (CFS) filter methods and wrapper feature selection methods on five widely-used prediction models and demonstrate how these models perform with or without feature selection to predict the number of bugs in five different open source Java software systems.
Our results show that wrappers outperform the CFS filter; they improve prediction accuracy by up to 33\% while eliminating more than half of the features.
We also observe that though the same feature selection method chooses different feature subsets in different projects, this subset always contains a mix of source code and change metrics.
\end{abstract}


\section{Introduction}
\label{intro}
In the field of machine learning, when there is a large number of features, determining the smallest subset that exhibits the strongest effect often decreases model complexity and increases prediction accuracy.
This process is called feature selection.\footnote{Feature selection is also known as variable selection, attribute selection, and variable subset selection.} There are two well-known types of feature selection methods: filters and wrappers.
Filters select features based on their relevance to the response variable independently of the prediction model.
Wrappers select features that increase the prediction accuracy of the model.
However, as with any design decision during the construction of a prediction model, one needs to evaluate different feature selection methods in order to choose one, and above all to assess whether it is needed or not.

While there has been extensive research on the impact of feature selection on prediction models in different domains, our investigation reveals that it is a rarely studied topic in the domain of bug prediction.
Few studies explore how feature selection affects the accuracy of classifying software entities into buggy or clean~\cite{Shiv13a}\cite{Gao2011a}\cite{Cata09b}\cite{Kris11a}\cite{Wang12a}\cite{Khos10a}\cite{Khos14a}\cite{Ghot17a}, but to the best of our knowledge no dedicated study exists on the impact of feature selection on the accuracy of predicting the number of bugs.
As a result of this research gap, researchers often overlook feature selection and provide their prediction models with all the metrics they have on a software project or in a dataset.
We argue that feature selection is a mandatory step in the bug prediction pipeline and its application might alter previous findings in the literature, especially when it comes to comparing different machine learning models or different software metrics.

In this paper we treat bug prediction as a regression problem where a bug predictor predicts the number of bugs in software entities as opposed to classifying software entities as buggy or clean.
We investigate the impact of filter and wrapper feature selection methods on the prediction accuracy of five machine learning models: K-Nearest Neighbour, Linear Regression, Multilayer Perceptron, Random Forest, and Support Vector Machine.
More specifically, we carry out an empirical study on five open source Java projects: Eclipse JDT Core, Eclipse PDE UI, Equinox, Lucene, and Mylyn to answer the following research questions:
\vspace{0.2cm}
\\\emph{RQ1: How does feature selection impact the prediction accuracy?}
Our results show that applying correlation-based feature selection (CFS) improves the prediction accuracy in 32\% of the experiments, degrades it in 24\%, and keeps it unchanged in the rest.
On the other hand, applying the wrapper feature selection method improves prediction accuracy by up to 33\% in 76\% of the experiments and never degrades it in any experiment. However, the impact of feature selection varies depending on the underlying machine learning model as different models vary in their sensitivity to noisy, redundant, and correlated features in the data.
We observe zero to negligible effect in the case of Random Forest models.
\vspace{0.2cm}
\\\emph{RQ2: Are wrapper feature selection methods better than filters?}
Wrapper feature selection methods are consistently either better than or similar to CFS.
Applying wrapper feature selection eliminates noisy and redundant features and keeps only relevant features for that specific project, increasing the prediction accuracy of the machine learning model.
\vspace{0.2cm}
\\\emph{RQ3: Do different methods choose different feature subsets?}
We realize there is no optimal feature subset that works for every project and feature selection should be applied separately for each new project.
We find that not only different methods choose different feature subsets on the same projects, but also the same feature selection method chooses different feature subsets for different projects.
Interestingly however, all selected feature subsets include a mix of change and source code metrics.
\vspace{0.2cm}

In summary, this paper makes the following contributions:
\begin{enumerate}
\item A detailed comparison between filter and wrapper feature selection methods in the context of bug prediction as a regression problem.
\item A detailed analysis on the impact of feature selection on five widely-used machine learning models in the literature.
\item A comparison between the selected features by different methods.
\end{enumerate}

The rest of this paper is organized as follows: In \autoref{background}, we give a technical background about feature selection in machine learning.
We motivate our work in \autoref{motivation}, and show how we are the first to study wrapper feature selection methods when predicting the number of bugs.
In \autoref{empirical}, we explain the experimental setup, discuss the results of our empirical study, and elaborate on the threats to validity of the results.
Finally, we discuss the related work in \autoref{relatedWork} showing how our findings are similar or different from the state of the art, then conclude this paper in \autoref{conclusions}.


\begin{figure}
\center{\includegraphics[width=1.0\linewidth]{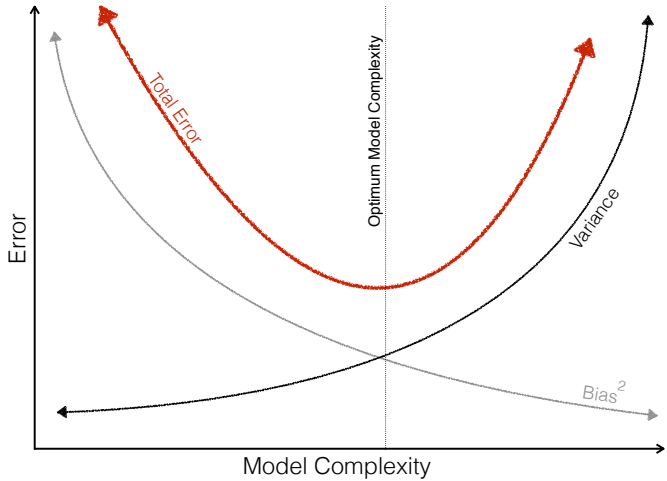}}
\caption{The relationship between model complexity and model error \cite{BiasVariance}.}
\label{fig:complexity}
\end{figure}

\section{Technical Background}
\label{background}
Trained on bug data and software metrics, a bug predictor is a machine learning model that predicts defective software entities using software metrics.
The software metrics are called the independent variables or the features.
The prediction itself is called the response variable or the dependent variable.
If the response variable is the absence/presence of bugs then bug prediction becomes a classification problem and the machine learning model is called a classifier.
If the response variable is the number of bugs in a software entity then bug prediction is a regression problem and the model is called a regressor.

Feature selection is an essential part in any machine learning process.
It aims at removing irrelevant and correlated features to achieve better accuracy, build faster models with stable performance, and reduce the cost of collecting features for later models.
Model error is known to be increased by both noise \cite{Atla11a} and feature multicollinearity \cite{Alle97c}.
Different feature selection algorithms eliminate this problem in different ways.
For instance, correlation based filter selection chooses features with high correlation with the response variable and low correlation with each other.

Also when we build a prediction model, we often favour less complex models over more complex ones due to the known relationship between model complexity and model error, as shown in Figure \autoref{fig:complexity}.
Feature selection algorithms try to reduce model complexity down to the sweet spot where the total error is minimal.
This point is called the optimum model complexity.
Model error is computed via the mean squared error (MSE) as: 
\\$MSE = \frac{1}{N} \sum_{i=1}^{N} (\hat{Y}_i - Y_i)^2$ \\where $\hat{Y}_i$ is the predicted value and $Y_i$ is the actual value.
MSE can be decomposed into model bias and model variance as: \\$MSE = Bias^2 + Variance + Irreducible Error $ \cite{Hast05a}

Bias is the difference between the average prediction of our model to the true unknown value we are trying to predict.
Variance is the variability of a model prediction for a given data point.
As can be seen in Figure \autoref{fig:complexity}, reducing model complexity increases the bias but decreases the variance.
Feature selection sacrifices a little bit of bias in order to reduce variance and, consequently, the overall MSE.

Every feature selection method consists of two parts: a search strategy and a scoring function.
The search strategy guides the addition or removal of features to the subset at hand and the scoring function evaluates the performance of that subset.
This process is repeated until no further improvement is observed.

\begin{table}
\scriptsize
\caption{The CK Metrics Suite \cite{Chid94a} and other object-oriented metrics included as the source code metrics in the bug prediction dataset \cite{DAmb10c}}
\begin{center}
\begin{tabular}{ll} 
{Metric Name}		& {Description}\\ \hline
CBO   & Coupling Between Objects \\[0.05cm] 
DIT   & Depth of Inheritance Tree \\[0.05cm] 
FanIn   & Number of classes that reference the class \\[0.05cm] 
FanOut   & Number of classes referenced by the class \\[0.05cm] 
LCOM   & Lack of Cohesion in Methods \\[0.05cm] 
NOC   & Number Of Children \\[0.05cm] 
NOA   & Number Of Attributes in the class \\[0.05cm] 
NOIA   & Number Of Inherited Attributes in the class \\[0.05cm] 
LOC   & Number of lines of code \\[0.05cm] 
NOM   & Number Of Methods \\[0.05cm] 
NOIM   & Number of Inherited Methods \\[0.05cm] 
NOPRA   & Number Of PRivate Atributes \\[0.05cm] 
NOPRM   & Number Of PRivate Methods \\[0.05cm] 
NOPA   & Number Of Public Atributes \\[0.05cm] 
NOPM   & Number Of Public Methods \\[0.05cm] 
RFC   & Response For Class \\[0.05cm] 
WMC   & Weighted Method Count  \\\hline

\label{tbl:sourceMetrics}
\end{tabular}
\end{center}
\end{table}

\begin{table}
\scriptsize
\caption{The change metrics proposed by Moser \etal \cite{Mose08a} included in the bug prediction dataset \cite{DAmb10c}}
\label{tbl:changeMetrics}
\begin{center}
\begin{tabularx}{0.48\textwidth}{lX} 
{Metric Name}		& {Description}\\ \hline
REVISIONS   & Number of reversions \\[0.05cm] 
BUGFIXES   & Number of bug fixes \\[0.05cm] 
REFACTORINGS   & Number Of Refactorings \\[0.05cm] 
AUTHORS   & Number of distinct authors that checked a file into the repository \\[0.05cm] 
LOC\_ADDED   & Sum over all revisions of the lines of code added to a file \\[0.05cm] 
MAX\_LOC\_ADDED   & Maximum number of lines of code added for all revisions \\[0.05cm] 
AVE\_LOC\_ADDED   & Average lines of code added per revision \\[0.05cm] 
LOC\_DELETED   & Sum over all revisions of the lines ofcode deleted from a file \\[0.05cm] 
MAX\_LOC\_DELETED   & Maximum number of lines of code deleted for all revisions \\[0.05cm] 
AVE\_LOC\_DELETED   & Average lines of code deleted per revision \\[0.05cm] 
CODECHURN   & Sum of (added lines of code - deleted lines of code) over all revisions \\[0.05cm] 
MAX\_CODECHURN   & Maximum CODECHURN for all revisions \\[0.05cm] 
AVE\_CODECHURN   & Average CODECHURN for all revisions \\[0.05cm] 
AGE   & Age of a file in weeks (counting backwards from a specific release) \\[0.05cm] 
WEIGHTED\_AGE   & Sum over age of a file in weeks times number of lines added during that week normalized by the total number of lines added to that file \\[0.05cm] \hline

\end{tabularx}
\end{center}
\end{table}

\section{Motivation}
\label{motivation}

In this section, we shortly discuss the importance of predicting the number of bugs in software entities.
Then, we highlight the impact of feature selection on bug prediction and particularly motivate the need for studying the wrapper methods.

\subsection{Regression vs Classification}
Most of the previous research treats bug prediction as a classification problem where software entities are classified as either buggy or clean, and there have been several studies on the impact of feature selection on defect classification models.
On the other hand, bug prediction as a regression problem is not well-studied, and the effect of feature selection on predicting the number of bugs is not well-understood.

Software bugs are not evenly distributed and tend to cluster \cite{Ostr04a}, and some software entities commonly have larger numbers of bugs compared to others.
Predicting the number of bugs in each entity provides valuable insights about the quality of these software entities~\cite{Osma16c}, which helps in prioritizing software entities to increase the efficiency of related development tasks such as testing and code reviewing~\cite{Khos03a}.
This is an important quality of a bug predictor especially for cost-aware bug prediction~\cite{Mend09b}\cite{Aris10a}\cite{Kame10a}\cite{Koba11a}\cite{Hata12a}.
In fact, predicting the number of bugs in software entities and then ordering these entities based on bug density is the most cost-effective option \cite{Osma17f}.

\subsection{Dimensionality Reduction}
\label{reduction}

When the dimensionality of data increases, distances grow more and alike between the vectors and it becomes harder to detect patterns in data~\cite{Bell60a}.
Feature selection not only eliminates the confounding effects of noise and feature multicollinearity, but also reduces the dimensionality of the data to improve accuracy.
However, feature selection does not seem to be considered as important as it should be in the field of bug prediction.
For instance, only 25 out of the 64 studied techniques in a recent research apply feature selection before training a machine learning model~\cite{Malh15a}.
Only 2 out of the 25 are applied to bug prediction as a regression problem.

\subsection{Filters vs Wrappers}
\label{filtersWrappers}
Feature selection methods are of two types: wrappers and filters \cite{Koha97a}.
With wrappers, the scoring function is the accuracy of the prediction model itself.
Wrappers look for the feature subset that works best with a specific machine learning model.
They are called wrappers because the machine learning algorithm is wrapped into the selection procedure.
With filters (\eg CFS, InfoGain, PCA), the scoring function is independent of the machine learning model.
They are called filters because the attribute set is filtered before the training phase.
Generally, filters are faster than wrappers but less powerful because wrappers address the fact that different learning algorithms can achieve best performance with different feature subsets.
In this paper we aim at finding whether there is actually a difference between filters and wrappers in bug prediction, and then quantifying this difference.

Wrappers are known to be computationally expensive.
They become a bottleneck when the size of a dataset (features + data items) becomes large.
However, this rarely happens in the bug prediction and bug prediction datasets tend to be relatively small.
This means that although wrappers are more resource intensive, they are easily applicable to bug prediction.
Nevertheless, our literature research yielded relatively few works that use wrappers for predicting number of bugs.

\section{Empirical Study}
\label{empirical}

In this section, we investigate the effect of feature selection on the accuracy of predicting the number of bugs in Java classes.
Specifically, we compare five widely-used machine learning models applied to five open source Java projects to answer the following research questions:
\\\emph{RQ1: How does feature selection impact the prediction accuracy?}
\\\emph{RQ2: Are wrapper feature selection methods better than filters?}
\\\emph{RQ3: Do different methods choose different feature subsets?}

\begin{table*}[h!]
\renewcommand{\arraystretch}{1.0}
\footnotesize
\caption{Details about the systems in the studied dataset, as reported by D'Ambros \etal~\cite{DAmb10c}}
\begin{center}
\begin{tabular}{llrrrr} 
				&		&				&			&				&\% classes with more \\
System 			&Release	&KLOC			&\#Classes  	& \% Buggy		&than one bug	\\ \hline

Eclipse JDT Core	&3.4		&$\approx 224 $	&997			& $\approx 20\%$	& $\approx 7\%$\\ 
			
Eclipse PDE UI		&3.4.1	&$\approx 40 $		&1,497		& $\approx 14\%$	 & $\approx 5\%$  \\ 
			
Equinox			&3.4		&$\approx 39 $		&324			& $\approx 40\%$   	& $\approx 15\%$ \\ 
			
Mylyn			&3.41	&$\approx 156$	&1,862		& $\approx 13\%$	& $\approx 4\%$    \\ 
			
Lucene			&2.4.0	&$\approx 146$	&691			& $\approx 9\%$  	& $\approx 3\%$\\ \hline
\label{tbl:dataset}
\end{tabular}
\end{center}
\end{table*}

\subsection{Experimental Setup}
\subsubsection*{Dataset}
We adopt the ``Bug Prediction Dataset" provided by D'Ambros \etal \cite{DAmb10c} which serves as a benchmark for bug prediction studies.
We choose this dataset because it is the only dataset that contains both source code and change metrics at the class level, in total 32 metrics listed in~\autoref{tbl:sourceMetrics} and \autoref{tbl:changeMetrics}; and also provides the number of post-release bugs as the response variable for five large open source Java systems listed in \autoref{tbl:dataset}.
The other dataset that has the number of bugs as a response variable comes from the PROMISE repository, but contains only 21 source code metrics \cite{Jure10a}.
 
\subsubsection*{Prediction Models}
We use Multi-Layer Perceptron (MLP), Random Forest (RF), Support Vector Machine (SVM), Linear Regression (LR), and an implementation of the k-nearest neighbour algorithm called IBK.
Each model represents a different category of statistical and machine learning models that is widely used in the bug prediction research~\cite{Malh15a}.

We use the correlation-based feature selection (CFS) method \cite{Hall00a}, the best \cite{Chal08b}\cite{Ghot17a} and the most commonly-used filter method in the literature \cite{Malh15a}.
For the wrapper feature selection method we use the corresponding wrapper applicable to each prediction model.
In other words, we use MLP wrapper for MLP, RF wrapper for RF, SVM wrapper for SVM, LR wrapper for LR, and IBK wrapper for IBK.
Every feature selection method also needs a search algorithm.
We use the \emph{Best First} search algorithm which searches the space of feature subsets using a greedy hill-climbing procedure with a backtracking facility.
We use this search algorithm because it returns the results in a reasonable amount of time while being exhaustive to a certain degree.
 
We use the Weka data mining tool \cite{Hall09} to build prediction models for each project in the dataset.
Following an empirical method similar to that of Hall and Holmes \cite{Hall03a}, we apply each prediction model to three feature sets: the full set, the subset chosen by CFS, and the subset chosen by the wrapper.
The prediction model is built and evaluated following the 10-fold cross validation procedure.
The wrapper feature selection is applied using a 5-fold cross validation on the training set of each fold, then the best feature set is used.
The CFS algorithm is applied to the whole training set of each fold.
Then the whole process is repeated 30 times.
We evaluate the predictions by means of the root mean squared error (RMSE).
In total, we have 25 experiments.
Each experiment corresponds to a specific project and a specific prediction model trained on the three feature sets.

We use the default hyperparameter (\ie configuration) values of Weka 3.8.0 for the used machine learning models.
Although hyperparameters can be tuned \cite{Tant16a}\cite{Osma17a}, we do not perform this optimization because we want to isolate the effect of feature selection.
Besides, Linear Regression does not have hyperparameters and the gained improvement of optimizing SVM and RF is negligible \cite{Tant16a}\cite{Osma17a}.

%

\subsection{Results}
\begin{figure*}
\center{\includegraphics[width=1.0\linewidth]{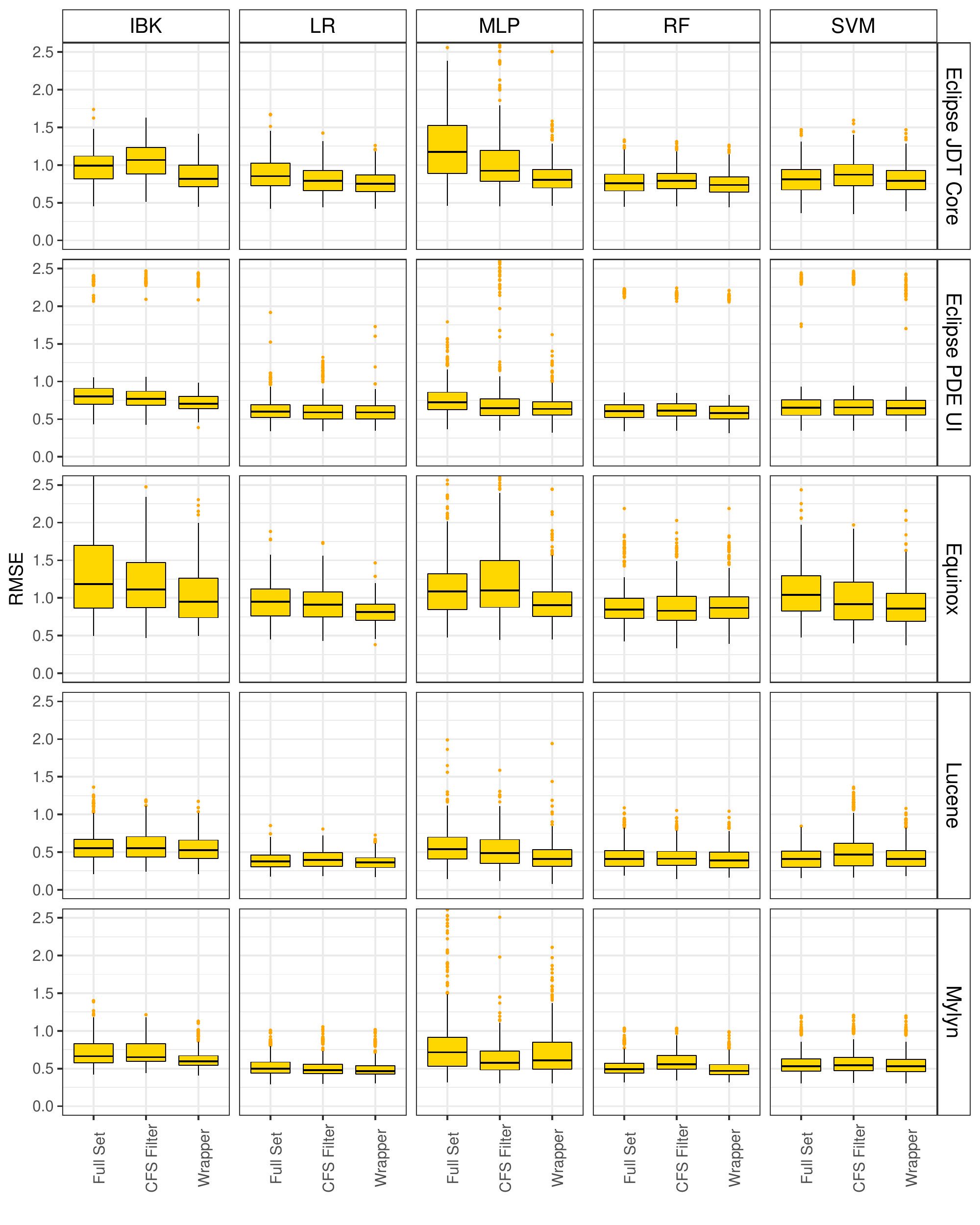}}
\caption{Boxplots of all the experiments in our empirical study.
The y-axis represents the root mean squared error (RMSE).
For each project/model, we examine three feature sets: the full set, the subset chosen by the CFS filter, and the subset chosen by the wrapper corresponding to the model.}
\label{fig:boxplots}
\end{figure*}

\autoref{fig:boxplots} shows standard box plots for the different RMSE values obtained by the different feature sets per prediction model per project.
Each box shows 50\% of that specific population.\footnote{By population we mean the RMSE values of a specific experiment with a specific feature set.
Each population consists of $10 \times 50 = 500$ data items (10-fold cross validation done 50 times)} We can see that the wrapper populations are almost always lower than the full set ones, have smaller boxes, and have fewer outliers.
This means that applying the wrapper gives better and more consistent predictions.
On the other hand, we cannot make any observations about applying CFS because the difference between the CFS populations and the full set populations are not apparent.

\begin{figure*}
\center{
\includegraphics[width=1.0\linewidth]{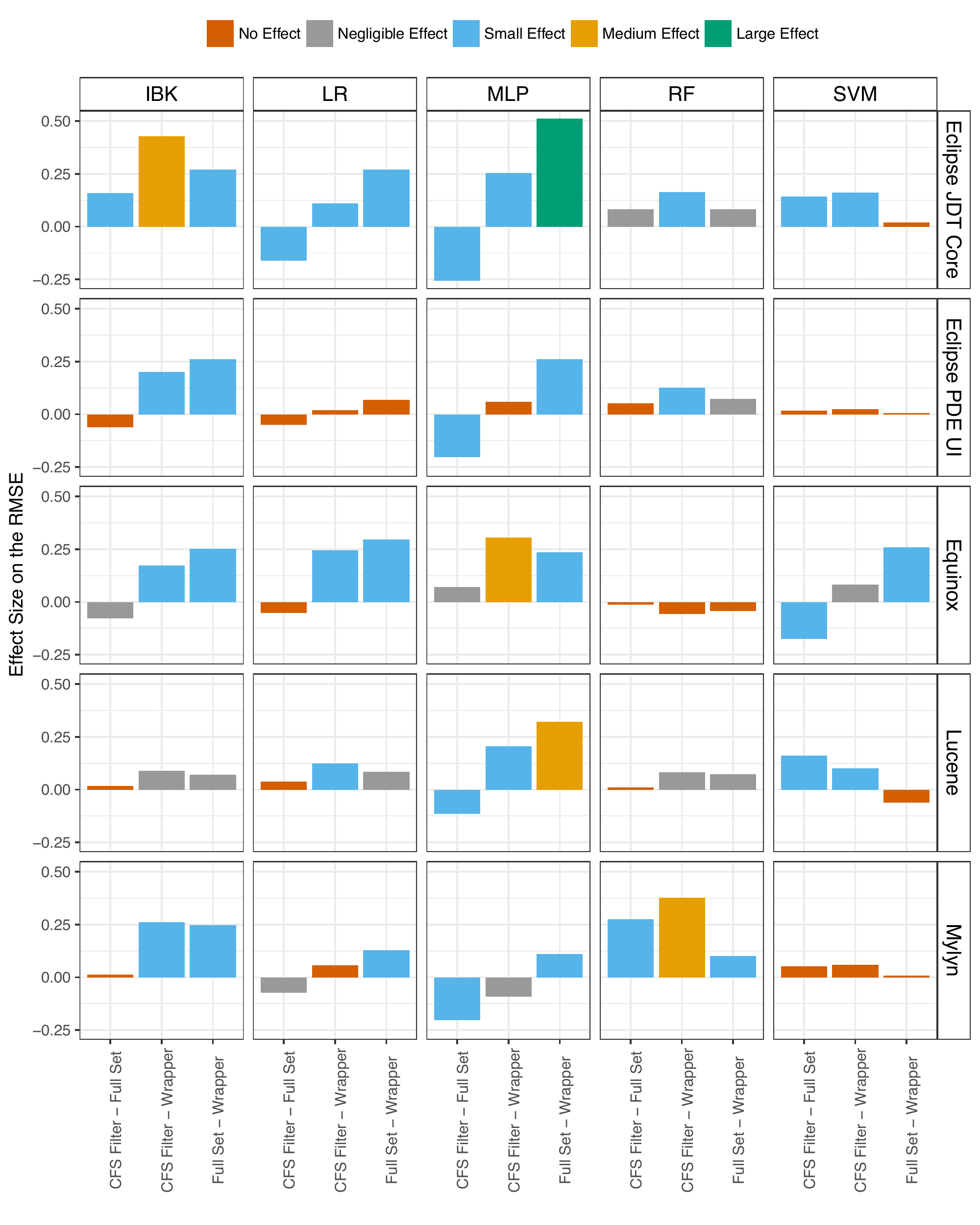}}
\caption{This figure shows the bar plots of the effect size of the Dunn post-hoc analysis, which is carried out at the 95\% confidence interval.
The x-axis indicates the pairwise comparison and the y-axis indicates the effect size.
The bars are color-coded.
If the bar is red, this means that the difference is not statistically significant.
Grey means that there is a statistical significant difference, but the effect is negligible.
Blue, golden, and green indicate a small, medium, and large statistically significant effect, respectively.}
\label{fig:effectSize}
\end{figure*}

While box plots are usually good to get an overview of the different populations and how they compare to each other, they do not provide any statistical evidence.
To get more concrete insights, we follow the two-stage statistical test: Kruskal-Wallis + Dunn post-hoc analysis, both at the 95\% confidence interval.
We apply the Kruskal-Wallis test on the results to determine whether different feature subsets have different prediction accuracies (\ie different RMSE).
Only when this test indicates that the populations are different, can we quantify such differences with a post-hoc analysis.
We perform Dunn post-hoc pairwise comparisons and analyze the effect size between each two populations.
\autoref{fig:effectSize} shows on the y-axis the detailed effect size between the two compared RMSE populations on the x-axis.
In this plot, there are two possible scenarios:
\begin{enumerate}
\item The Kruskal-Wallis test indicates that there is no statistical difference between the populations.
Then all the bars are red to show that there is no effect between any two populations.
\item The Kruskal-Wallis test indicates a statistically significant difference between the populations.
Then the color of the bars encode the pairwise effect size.
Red means no difference and the two populations are equivalent.
Grey means that there is a significant difference but can be ignored due to the negligible effect size.
Blue, golden, and green mean small, medium, and large effect size respectively.
\end{enumerate}

To see how feature selection methods impact the prediction accuracy (\emph{RQ1}), we compare the RMSE values obtained by applying CFS and wrappers with those obtained by the full feature set.
We observe that the RMSE value obtained by the CFS feature subset is statistically lower than the full set in 8 experiments (32\%),\footnote{negative non-red effect size in \autoref{fig:effectSize}} statistically higher in other 6 experiments (24\%),\footnote{positive non-red effect size in \autoref{fig:effectSize}} and statistically equivalent in 11 experiments (44\%).\footnote{red effect size in \autoref{fig:effectSize}} Although CFS can decrease the RMSE by 24\% on average (MLP with Mylyn), it can increase it by up to 24\% (SVM with Lucene).
We also notice that applying CFS is not consistent within experiments using the same model.
It does not always improve, or always degrade, or always retain the performance of any model throughout the experiments.
We conclude that CFS is unreliable and gives unstable results.
Furthermore, even when CFS reduces the RMSE, the effect size is at most small.

On the other hand, the RMSE value of the wrapper feature subset is statistically lower than that of the full set in 19 experiments (76\%)  and statistically equivalent in the rest.
Applying the wrapper feature selection method can decrease RMSE of a model by up to 33\% (MLP with Eclipse JDT).
We also observe that the impact of the wrapper feature selection method on the accuracy is different from one model to another.
It has a non-negligible improvement on the prediction accuracy of IBK, LR, MLP, RF, and SVM in 80\%, 60\%, 100\%, 20\%, and 20\% of the experiments, respectively.
This is due to the fact that different machine learning models are different in their robustness against noise and multicollinearity.
MLP, IBK, and LR were improved significantly almost always in our experiments.
On the other hand, SVM and RF were not improved as often, because they are known to be resistant to noise, especially when the number of features is not too high.
RF is an ensemble of decision trees created by using bootstrap samples of the training data and random feature selection in tree induction \cite{Brei01a}.
This gives RF the ability to work well with high-dimensional data and sift the noise away.
The SVM algorithm is also designed to operate in a high-dimensional feature space and can automatically select relevant features \cite{Hsu03a}.
In fact, this might be the reason behind the proven record of Random Forest and Support Vector Machine in bug prediction \cite{Guo04a}\cite{Elis08a}.

\begin{table}[]
\renewcommand{\arraystretch}{1.2}
\normalsize
\caption{The level of agreement between different feature selection methods in each project}
\begin{center}
\begin{tabular}{lll} 
Project 			& $k$ 	&Agreement \\ \hline
Eclipse JDT Core	&0.18	& Slight\\ 
			
Eclipse PDE UI		&0.17	 & Slight   \\ 
			
Equinox			&0.40	  & Fair  \\ 
			
Mylyn			&0.08	 & Slight   \\ 
			
Lucene			&0.18	& Slight  \\ \hline
\label{tbl:kappaOfProjects}
\end{tabular}
\end{center}
\end{table}

\begin{table}[]
\renewcommand{\arraystretch}{1.2}
\normalsize
\caption{The level of agreement between the feature subsets selected by each method over all projects}
\begin{center}
\begin{tabular}{lll} 
Feature Selection Method 			& $k$ 	&Agreement \\ \hline
CFS								&0.23	& Fair\\ 
			
IBK Wrapper						&0.26	 & Fair   \\ 
			
LR Wrapper						&0.16	  & Slight  \\ 
			
MLP Wrapper						&0.04	 & Slight   \\ 

RF Wrapper						&0.04	 & Slight   \\ 
			
SVM Wrapper						&-0.01	& Poor  \\ \hline

\label{tbl:kappaOfMethods}
\end{tabular}
\end{center}
\end{table}

The wrapper method is statistically better than CFS in 18 experiments, statistically equivalent in 6 experiments, and worse in one experiment, but with a negligible effect size.
These results along with the fact that CFS sometimes increases the RMSE, clearly show that the wrapper selection method is a better choice than CFS (\emph{RQ2}).

\begin{figure*}[]
     \begin{center}
        \subfigure[]{%
            \label{fig:selectedFeaturesGrid}
            \includegraphics[width=0.9\textwidth]{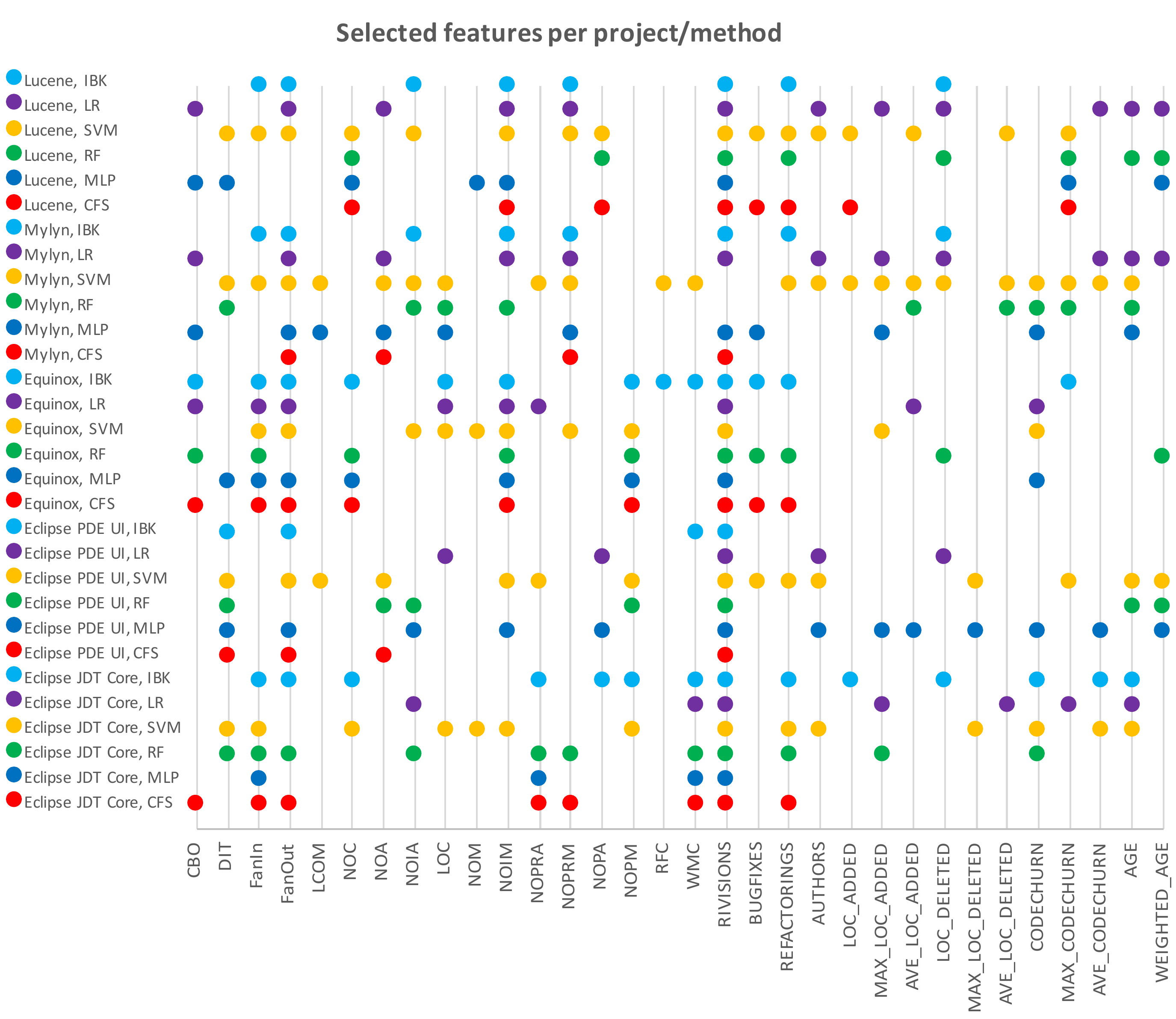}}

        \subfigure[]{%
            \label{fig:selectedFeaturesCounts}
            \includegraphics[width=0.9\textwidth]{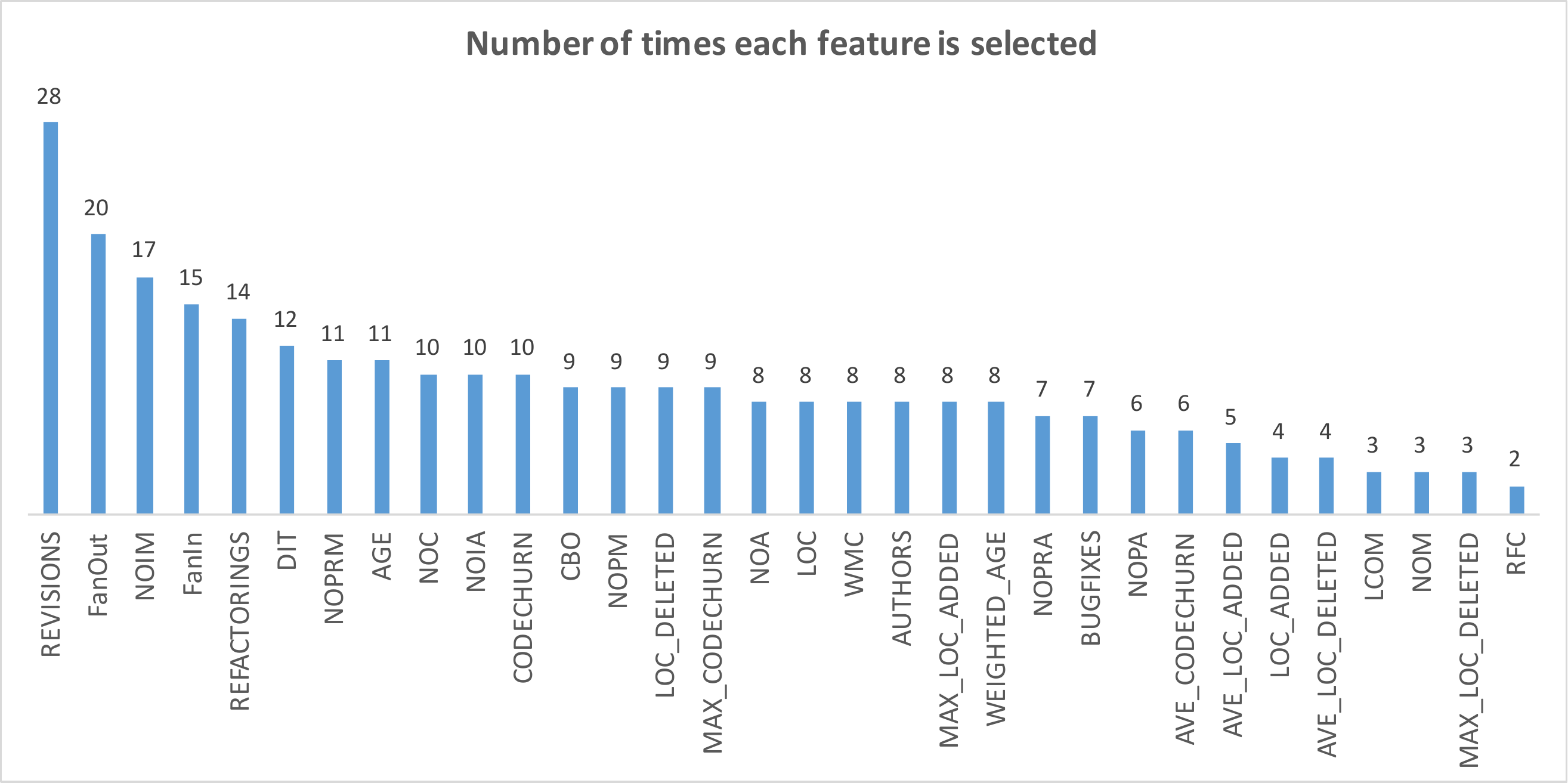}}
    \end{center}
\end{figure*}
\begin{figure*}[]
     \begin{center}

 \subfigure[]{%
            \label{fig:selectedFeaturesPerProject}
            \includegraphics[width=1\textwidth]{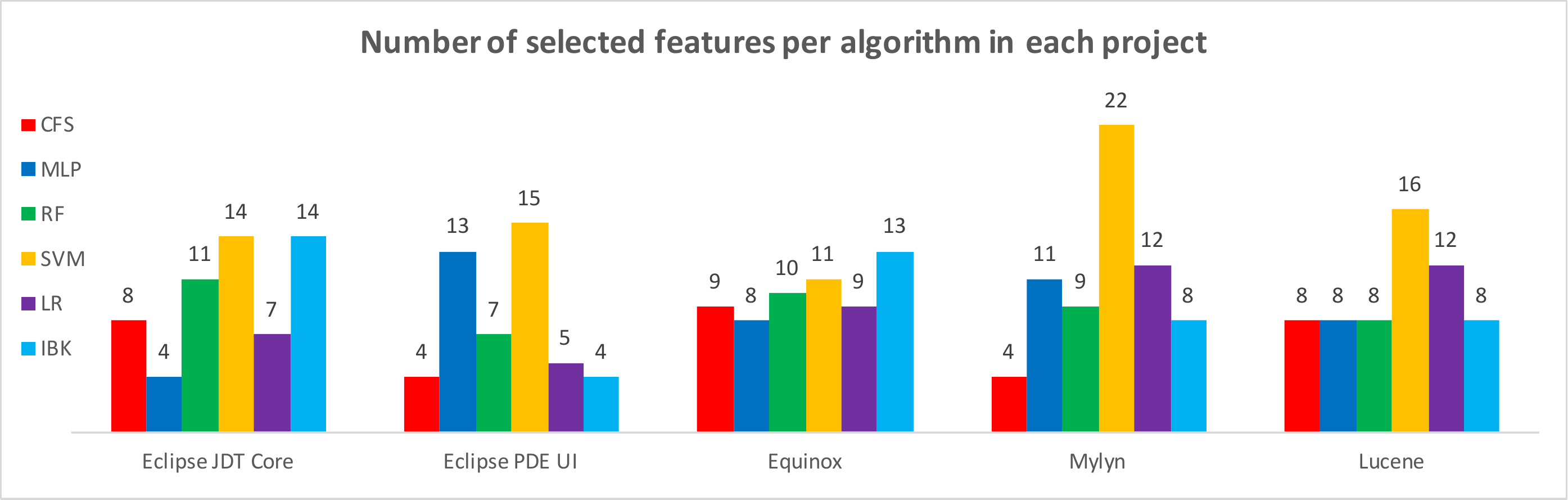}}

    \end{center}
    \caption{Subfigure (a) shows the features selected by each method using the whole data of each project.
Subfigure (b) shows the number of times each feature is selected out of the 30 (1 CFS feature set + 5 wrapper sets per project).
The more times a feature is selected the more important it is for making accurate predictions.
Subfigure (c) shows how different selection methods vary in the number of selected features.
Details about the features (metrics) are in \autoref{tbl:sourceMetrics} and \autoref{tbl:changeMetrics}}
   \label{fig:selectedFeatures}
\end{figure*}

\autoref{fig:selectedFeatures} shows the details about the features selected by each method using the whole data of each project in the dataset.
To answer the third research question (\emph{RQ3}), we use the Fleiss' kappa statistical measure \cite{Flei71a} to evaluate the level of agreement between the different feature selection methods for each project and the level of agreement of each feature selection method over the different projects. The Fleiss' kappa value, called $k$, is interpreted as follows:
\\$k\leq0 \implies$ poor agreement
\\$0.01 <k \leq 0.20 \implies$ Slight agreement
\\$0.21 <k \leq 0.40 \implies$ Fair agreement
\\$0.41 <k \leq 0.60 \implies$ Moderate agreement
\\$0.61 <k \leq 0.80 \implies$ Substantial agreement
\\$0.81 <k \leq 1.00 \implies$ Almost perfect agreement


Figure \autoref{fig:selectedFeaturesGrid} shows that different methods choose different features in each project.
The level of agreement between the different methods is \emph{slight} in four projects and \emph{fair} in only one, as detailed in \autoref{tbl:kappaOfProjects}.
Also the same method chooses different features in different projects.
\autoref{tbl:kappaOfMethods} shows that the level of agreement between the feature subsets selected by the same method in different projects is at most \emph{fair}.
However, there exists some agreement on some features.
Figure \autoref{fig:selectedFeaturesCounts} shows that \emph{REVISIONS, FanOut, NOIM,} and \emph{FanIn} are chosen most often.
\emph{REVISIONS} in particular is chosen by all methods almost all the time.
It is selected in 28 out of 30 feature subsets,\footnote{For each one of the 5 projects in the dataset, there are 6 feature subsets: 1 CFS subset and 5 wrapper subsets.} meaning that it has a high predictive power.
On the other hand, \emph{RFC, MAX\_LOC\_DELETED, NOM}, and \emph{LCOM} are picked the least number of times, which means they have little to no predictive power.

The number of discarded features varies from 10 to 28 features out of 32 in total, as detailed in Figure \autoref{fig:selectedFeaturesPerProject}.
This means that most of the features can be removed while enhancing (in 48\% of the experiments), or at least retaining (in 52\% of the experiments), the prediction accuracy.

Another important observation is that no feature subset contains only source code metrics or only change metrics, but always a mix of both.
This means that no category of metrics (\ie change and source code) alone is good for predicting the number of bugs, but they should be combined to achieve better performance.
Previous studies show that change metrics are better than source code metrics in bug prediction and combining the two sets either does not bring any benefit \cite{Mose08a} or hinders the performance \cite{Aris10a}.
However, these studies either did not employ feature selection at all \cite{Mose08a} or employed only CFS \cite{Aris10a}.

\subsection{Threats to Validity}

Although we use a well-known benchmark as our dataset, the quality of our results is very dependent on the quality of that dataset.
Also our dependence on WEKA for building the machine learning models, makes the quality of the models dependent solely on the quality of the WEKA implementation itself.
Since we are dealing with machine learning, the risk of overfitting always exists.
In general, we mitigate this risk by repeating each feature selection/training/testing process 50 times.
Also for the wrapper method, we apply the train-test-validate (TTV) approach to mitigate any overfitting bias in favour of wrappers.\footnote{The train-test-validate is not applicable to CFS due to how CFS operates} 

In our study, the dataset contains metrics only from open source software systems.
It is hard to generalize the results to all systems due to the differences between industrial and open source projects.
Also all the systems in the dataset are developed in Java and the findings might not generalize to systems developed in other languages.
Another important threat to external validity comes from the chosen prediction models.
Although we try to pick a representative set of prediction models, our findings might not generalize to other machine learning models such as hybrid and genetic ones.

Machine learning methods depend on the underlying data.
Although our findings are statistically strong, they may not generalize to all open source Java systems.
Practitioners and researchers are encouraged to run comparative studies, similar to ours, before making decisions regarding the machine learning techniques employed for bug prediction.

\section{Related Work}
\label{relatedWork}

In a recent systematic literature review of the machine learning techniques in the bug prediction field~\cite{Malh15a}, Most studies (60\%) do not apply feature selection at all.
Our investigation shows that researchers in bug prediction often undermine the importance of feature selection.

Among the studies that apply feature selection, CFS \cite{Hall00a} is the most commonly used \cite{Malh15a}.
Many studies employ the CFS filter technique before training machine learning models \cite{Elis08a}\cite{Kaur08a}\cite{Aris10a} \cite{DeCa10a}\cite{Malh11a}\cite{Zimm09a}\cite{Twal11a}\cite{Okut14a}.
We also apply CFS as the baseline filter technique and compare it to wrapper techniques to show that wrappers outperform this filter in most cases.

Most studies that apply feature selection actually apply filter techniques like principal component analysis (PCA) \cite{Khos06a}\cite{DAmb10c}\cite{Naga06a}\cite{Kanm07a}\cite{Sing10a}, consistency based selection (CBS) \cite{Zimm09a}, and InfoGain \cite{Menz07a}\cite{Turh07a}\cite{Sing09a}\cite{Misi11a}.
Very few studies apply wrapper feature selection techniques \cite{Cahi13a}\cite{Gond08a} when classifying software entities as buggy or clean (classification).
However, to the best of our knowledge, there is no study that applies wrapper techniques when predicting the number of bugs in software entities (regression).
In our study we compare the CFS filter with different wrappers while treating bug prediction as a regression problem.

Shivaji \etal \cite{Shiv13a} study the impact of multiple feature selection techniques (filters and wrappers) on the performance of Na\"ive Bayes and Support Vector Machines when classifying code changes as buggy or clean.
They report a significant enhancement in the accuracy measures of both classifiers when feature selection is applied.
However, Shivaji \etal consider \emph{``everything in the source code separated by whitespace or a semicolon"} as a feature \cite{Shiv13a}.
This includes variable names, method names, keywords, comment words, \etc
They end up with a staggering number of features ranging from 6,127 to 41,942 features.
The vast majority of features in the initial feature set are bound to be irrelevant, hence the results of their study are exaggerated and cannot be generalized.

Challagulla \etal \cite{Chal08b} report that performing principal component analysis before training the models does not result in a significant improvement of the performance of the bug predictors, while correlation-based feature selection (CFS) and consistency-based subset evaluation (CBS) actually decrease the prediction error of the machine learning models.
We actually report different results when applying CFS to regressors.

Khoshgoftaar \etal \cite{Khos10a} combine filter feature selection and data sampling to achieve better classification accuracy on the package level.
They report an improvement of around 2\% to the area under the receiver operator characteristic curve (AUC-ROC) when the filters are applied to sampled data rather than the whole data.
Then Khoshgoftaar \etal \cite{Khos14a} improve the performance by repeatedly applying the sampling and filtering several times then aggregating the results.

In their defect-proneness prediction framework, Song \etal \cite{Song11a} included a special step for feature selection.
However, their framework treats bug prediction as a classification problem.
In this paper we study the effect of wrapper feature selection methods on bug prediction as a regression problem.

Gao \etal \cite{Gao2011a} studied seven filter feature selection techniques.
They report that the studied classification models were either improved or remained unchanged while 85\% of the original features were eliminated.
Krishnan \etal \cite{Kris11a} analyze whether change metrics remain good predictors during the evolution of Eclipse.
In their analysis, they use J48 decision tree as the machine learning algorithm and refer to the top five features as the set of prominent predictors, then study the consistency of this set over the consecutive versions of Eclipse.
They report that there is a small subset of change metrics that is consistently good at classifying software entities as buggy or clean across products and revisions.
Wang \etal \cite{Wang12a} study the effect of removing redundant or useless features from the PROMISE dataset.\footnote{\url{http://openscience.us/repo/}}
They report that feature selection improves classification accuracy.
Catal and Diri \cite{Cata09b} explore which machine learning algorithm performs best before and after applying feature reduction.
In these studies, researchers treat bug prediction as a classification problem while we study the effect of feature selection on bug prediction as a regression problem.

Turhan and Bener \cite{Turh09a} argue that Na\"ive Bayes assumes the independence and the equal importance of features.
These assumptions are not true in the context of bug prediction.
Yet, Na\"ive Bayes is one of the best classifiers \cite{Menz07a}.
They empirically show that the independence assumption is not harmful for defect prediction using Na\"ive Bayes and assigning weights to features increases the performance and removes the need for feature selection.
They conclude that \emph{``either weighted Naive Bayes or pre-processing data with PCA may produce better results to locate software bugs in source code"}\cite{Turh09a}.

We confirm that most of the existing literature treats bug prediction as a classification problem, and that studying the number of bugs is neglected in the field.
For instance, Ostrand \etal \cite{Ostr05a} use negative binomial regression (NBR) to predict the number of bugs in software modules.
They report that NBR fits the bug prediction problem and demonstrates high accuracy.
Janes \etal \cite{Jane06a} compare three count models to predict the number of bugs and find out that zero-inflated NBR performs better than Poisson regression and NBR.
Rathore and Kumar \cite{Rath16a} investigate six different fault prediction models for predicting the number of bugs and show that count models (\ie NBR and Zero-Inflated Poisson regression) underperform compared to linear regression, decision tree regression, genetic programming and multilayer perceptron.
Graves \etal \cite{Grav00a} build a generalized linear regression model to predict the number of bugs based on the various change metrics.
Gao and Khoshgoftaar \cite{Gao07a} compare the performance of several count models (\eg Poisson regression) and show that Hurdle Poisson regression has the most accurate predictions.
Nevertheless, these studies do not apply feature selection and their results should be reassessed in the light of our findings.

\section{Conclusions}
\label{conclusions}
Generalizing bug prediction findings is hard.
Software projects have different teams, cultures, frameworks, and architectures.
Consequently, software metrics have different correlations with the number of bugs in different projects.
These correlations can be captured differently by distinct prediction models.
We argue that wrapper feature selection methods fit this problem best because they not only choose features relevant to the response variable but also to the prediction model itself.
Indeed, our results show that wrapper feature selection is always better than CFS and improves the performance of a model (by up to 47\%) while eliminating most of the features (up to 87\%).
Our results also reveal that there is little agreement on the selected features.
Even the same method chooses different features in different projects.
We cannot generalize what feature subset to use, but we can recommend combining both change and source code metrics and letting the wrapper feature selection method choose the right subset.

In the future, we plan to investigate how the selected feature subset changes with the evolution of a certain project from the size and selection perspectives.
Lastly, while carrying out this research we realized that datasets providing the number of bugs as the response variable are scarce, which could be a hurdle to studies predicting the number of bugs in software systems.
We therefore encourage the community to publish such defect datasets more frequently.
\section*{Acknowledgments}
The authors gratefully acknowledge the funding of the Swiss National Science Foundations for the project ``Agile Software Analysis" (SNF project No. 200020\_162352, Jan 1, 2016 - Dec. 30, 2018).\footnote{\url{http://p3.snf.ch/Project-162352}}
\bibliographystyle{IEEEtran}
\bibliography{scg}

\begin{thebibliography}{10}
\providecommand{\url}[1]{#1}
\csname url@samestyle\endcsname
\providecommand{\newblock}{\relax}
\providecommand{\bibinfo}[2]{#2}
\providecommand{\BIBentrySTDinterwordspacing}{\spaceskip=0pt\relax}
\providecommand{\BIBentryALTinterwordstretchfactor}{4}
\providecommand{\BIBentryALTinterwordspacing}{\spaceskip=\fontdimen2\font plus
\BIBentryALTinterwordstretchfactor\fontdimen3\font minus
  \fontdimen4\font\relax}
\providecommand{\BIBforeignlanguage}[2]{{%
\expandafter\ifx\csname l@#1\endcsname\relax
\typeout{** WARNING: IEEEtran.bst: No hyphenation pattern has been}%
\typeout{** loaded for the language `#1'. Using the pattern for}%
\typeout{** the default language instead.}%
\else
\language=\csname l@#1\endcsname
\fi
#2}}
\providecommand{\BIBdecl}{\relax}
\BIBdecl

\bibitem{Shiv13a}
S.~Shivaji, E.~J. Whitehead, R.~Akella, and S.~Kim, ``Reducing features to
  improve code change-based bug prediction,'' \emph{IEEE Transactions on
  Software Engineering}, vol.~39, no.~4, pp. 552--569, 2013.

\bibitem{Gao2011a}
K.~Gao, T.~M. Khoshgoftaar, H.~Wang, and N.~Seliya, ``Choosing software metrics
  for defect prediction: an investigation on feature selection techniques,''
  \emph{Software: Practice and Experience}, vol.~41, no.~5, pp. 579--606, 2011.

\bibitem{Cata09b}
C.~Catal and B.~Diri, ``Investigating the effect of dataset size, metrics sets,
  and feature selection techniques on software fault prediction problem,''
  \emph{Information Sciences}, vol. 179, no.~8, pp. 1040--1058, 2009.

\bibitem{Kris11a}
S.~Krishnan, C.~Strasburg, R.~R. Lutz, and K.~Go{\v{s}}eva-Popstojanova, ``Are
  change metrics good predictors for an evolving software product line?'' in
  \emph{Proceedings of the 7th International Conference on Predictive Models in
  Software Engineering}.\hskip 1em plus 0.5em minus 0.4em\relax ACM, 2011,
  p.~7.

\bibitem{Wang12a}
P.~Wang, C.~Jin, and S.-W. Jin, ``Software defect prediction scheme based on
  feature selection,'' in \emph{Information Science and Engineering (ISISE),
  2012 International Symposium on}.\hskip 1em plus 0.5em minus 0.4em\relax
  IEEE, 2012, pp. 477--480.

\bibitem{Khos10a}
T.~M. Khoshgoftaar, K.~Gao, and N.~Seliya, ``Attribute selection and imbalanced
  data: Problems in software defect prediction,'' in \emph{2010 22nd IEEE
  International Conference on Tools with Artificial Intelligence},
  vol.~1.\hskip 1em plus 0.5em minus 0.4em\relax IEEE, 2010, pp. 137--144.

\bibitem{Khos14a}
\BIBentryALTinterwordspacing
T.~M. Khoshgoftaar, K.~Gao, A.~Napolitano, and R.~Wald, ``A comparative study
  of iterative and non-iterative feature selection techniques for software
  defect prediction,'' \emph{Information Systems Frontiers}, vol.~16, no.~5,
  pp. 801--822, 2014. [Online]. Available:
  \url{http://dx.doi.org/10.1007/s10796-013-9430-0}
\BIBentrySTDinterwordspacing

\bibitem{Ghot17a}
\BIBentryALTinterwordspacing
B.~Ghotra, S.~Mcintosh, and A.~E. Hassan, ``A large-scale study of the impact
  of feature selection techniques on defect classification models,'' in
  \emph{Proceedings of the 14th International Conference on Mining Software
  Repositories}, ser. MSR '17.\hskip 1em plus 0.5em minus 0.4em\relax
  Piscataway, NJ, USA: IEEE Press, 2017, pp. 146--157. [Online]. Available:
  \url{https://doi.org/10.1109/MSR.2017.18}
\BIBentrySTDinterwordspacing

\bibitem{BiasVariance}
\BIBentryALTinterwordspacing
``Understanding the bias-variance tradeoff,'' accessed June 9, 2016,
  http://scott.fortmann-roe.com/docs/BiasVariance.html. [Online]. Available:
  \url{http://scott.fortmann-roe.com/docs/BiasVariance.html}
\BIBentrySTDinterwordspacing

\bibitem{Atla11a}
A.~Atla, R.~Tada, V.~Sheng, and N.~Singireddy, ``Sensitivity of different
  machine learning algorithms to noise,'' \emph{Journal of Computing Sciences
  in Colleges}, vol.~26, no.~5, pp. 96--103, 2011.

\bibitem{Alle97c}
M.~P. Allen, ``The problem of multicollinearity,'' \emph{Understanding
  Regression Analysis}, pp. 176--180, 1997.

\bibitem{Hast05a}
T.~Hastie, R.~Tibshirani, J.~Friedman, and J.~Franklin, ``The elements of
  statistical learning: data mining, inference and prediction,'' \emph{The
  Mathematical Intelligencer}, vol.~27, no.~2, pp. 83--85, 2005.

\bibitem{Chid94a}
\BIBentryALTinterwordspacing
S.~R. Chidamber and C.~F. Kemerer, ``A metrics suite for object oriented
  design,'' \emph{IEEE Transactions on Software Engineering}, vol.~20, no.~6,
  pp. 476--493, Jun. 1994. [Online]. Available:
  \url{http://dx.doi.org/10.1109/32.295895}
\BIBentrySTDinterwordspacing

\bibitem{DAmb10c}
M.~D'Ambros, M.~Lanza, and R.~Robbes, ``An extensive comparison of bug
  prediction approaches,'' in \emph{Proceedings of MSR 2010 (7th IEEE Working
  Conference on Mining Software Repositories)}.\hskip 1em plus 0.5em minus
  0.4em\relax IEEE CS Press, 2010, pp. 31--40.

\bibitem{Mose08a}
\BIBentryALTinterwordspacing
R.~Moser, W.~Pedrycz, and G.~Succi, ``A comparative analysis of the efficiency
  of change metrics and static code attributes for defect prediction,'' in
  \emph{Proceedings of the 30th International Conference on Software
  Engineering}, ser. ICSE '08.\hskip 1em plus 0.5em minus 0.4em\relax New York,
  NY, USA: ACM, 2008, pp. 181--190. [Online]. Available:
  \url{http://doi.acm.org/10.1145/1368088.1368114}
\BIBentrySTDinterwordspacing

\bibitem{Ostr04a}
T.~J. Ostrand, E.~J. Weyuker, and R.~M. Bell, ``Where the bugs are,'' in
  \emph{ACM SIGSOFT Software Engineering Notes}, vol.~29.\hskip 1em plus 0.5em
  minus 0.4em\relax ACM, 2004, pp. 86--96.

\bibitem{Osma16c}
\BIBentryALTinterwordspacing
H.~Osman, ``On the non-generalizability in bug prediction,'' in \emph{Post
  Proceedings of the Ninth Seminar on Advanced Techniques and Tools for
  Software Evolution (SATToSE 2016)}, 2016. [Online]. Available:
  \url{http://ceur-ws.org/Vol-1791/paper-03.pdf}
\BIBentrySTDinterwordspacing

\bibitem{Khos03a}
T.~M. Khoshgoftaar and E.~B. Allen, ``Ordering fault-prone software modules,''
  \emph{Software Quality Journal}, vol.~11, no.~1, pp. 19--37, 2003.

\bibitem{Mend09b}
\BIBentryALTinterwordspacing
T.~Mende and R.~Koschke, ``Revisiting the evaluation of defect prediction
  models,'' in \emph{Proceedings of the 5th International Conference on
  Predictor Models in Software Engineering}, ser. PROMISE '09.\hskip 1em plus
  0.5em minus 0.4em\relax New York, NY, USA: ACM, 2009, pp. 7:1--7:10.
  [Online]. Available: \url{http://doi.acm.org/10.1145/1540438.1540448}
\BIBentrySTDinterwordspacing

\bibitem{Aris10a}
\BIBentryALTinterwordspacing
E.~Arisholm, L.~C. Briand, and E.~B. Johannessen, ``A systematic and
  comprehensive investigation of methods to build and evaluate fault prediction
  models,'' \emph{J. Syst. Softw.}, vol.~83, no.~1, pp. 2--17, Jan. 2010.
  [Online]. Available: \url{http://dx.doi.org/10.1016/j.jss.2009.06.055}
\BIBentrySTDinterwordspacing

\bibitem{Kame10a}
Y.~Kamei, S.~Matsumoto, A.~Monden, K.-i. Matsumoto, B.~Adams, and A.~Hassan,
  ``Revisiting common bug prediction findings using effort-aware models,'' in
  \emph{Software Maintenance (ICSM), 2010 IEEE International Conference on},
  Sep. 2010, pp. 1--10.

\bibitem{Koba11a}
\BIBentryALTinterwordspacing
K.~Kobayashi, A.~Matsuo, K.~Inoue, Y.~Hayase, M.~Kamimura, and T.~Yoshino,
  ``{ImpactScale}: Quantifying change impact to predict faults in large
  software systems,'' in \emph{Proceedings of the 2011 27th IEEE International
  Conference on Software Maintenance}, ser. ICSM '11.\hskip 1em plus 0.5em
  minus 0.4em\relax Washington, DC, USA: IEEE Computer Society, 2011, pp.
  43--52. [Online]. Available:
  \url{http://dx.doi.org/10.1109/ICSM.2011.6080771}
\BIBentrySTDinterwordspacing

\bibitem{Hata12a}
\BIBentryALTinterwordspacing
H.~Hata, O.~Mizuno, and T.~Kikuno, ``Bug prediction based on fine-grained
  module histories,'' in \emph{Proceedings of the 34th International Conference
  on Software Engineering}, ser. ICSE '12.\hskip 1em plus 0.5em minus
  0.4em\relax Piscataway, NJ, USA: IEEE Press, 2012, pp. 200--210. [Online].
  Available: \url{http://dl.acm.org/citation.cfm?id=2337223.2337247}
\BIBentrySTDinterwordspacing

\bibitem{Osma17f}
\BIBentryALTinterwordspacing
H.~Osman, M.~Ghafari, O.~Nierstrasz, and M.~Lungu, ``An extensive analysis of
  efficient bug prediction configurations,'' in \emph{Proceedings of the 13th
  International Conference on Predictive Models and Data Analytics in Software
  Engineering}, ser. PROMISE.\hskip 1em plus 0.5em minus 0.4em\relax New York,
  NY, USA: ACM, 2017, pp. 107--116. [Online]. Available:
  \url{http://doi.acm.org/10.1145/3127005.3127017}
\BIBentrySTDinterwordspacing

\bibitem{Bell60a}
R.~Bellman, \emph{Adaptive control processes: A guided tour.}\hskip 1em plus
  0.5em minus 0.4em\relax Princeton University Press, St Martin's Press, 1960.

\bibitem{Malh15a}
R.~Malhotra, ``A systematic review of machine learning techniques for software
  fault prediction,'' \emph{Applied Soft Computing}, vol.~27, pp. 504--518,
  2015.

\bibitem{Koha97a}
R.~Kohavi and G.~H. John, ``Wrappers for feature subset selection,''
  \emph{Artificial intelligence}, vol.~97, no.~1, pp. 273--324, 1997.

\bibitem{Jure10a}
\BIBentryALTinterwordspacing
M.~Jureczko and L.~Madeyski, ``Towards identifying software project clusters
  with regard to defect prediction,'' in \emph{Proceedings of the 6th
  International Conference on Predictive Models in Software Engineering}, ser.
  PROMISE '10.\hskip 1em plus 0.5em minus 0.4em\relax New York, NY, USA: ACM,
  2010, pp. 9:1--9:10. [Online]. Available:
  \url{http://doi.acm.org/10.1145/1868328.1868342}
\BIBentrySTDinterwordspacing

\bibitem{Hall00a}
M.~A. Hall, ``Correlation-based feature selection for discrete and numeric
  class machine learning,'' in \emph{Proceedings of the Seventeenth
  International Conference on Machine Learning}.\hskip 1em plus 0.5em minus
  0.4em\relax Morgan Kaufmann Publishers Inc., 2000, pp. 359--366.

\bibitem{Chal08b}
V.~U.~B. Challagulla, F.~B. Bastani, I.-L. Yen, and R.~A. Paul, ``Empirical
  assessment of machine learning based software defect prediction techniques,''
  \emph{International Journal on Artificial Intelligence Tools}, vol.~17,
  no.~02, pp. 389--400, 2008.

\bibitem{Hall09}
M.~Hall, E.~Frank, G.~Holmes, B.~Pfahringer, P.~Reutemann, and I.~H. Witten,
  ``The weka data mining software: an update,'' \emph{ACM SIGKDD explorations
  newsletter}, vol.~11, no.~1, pp. 10--18, 2009.

\bibitem{Hall03a}
M.~A. Hall and G.~Holmes, ``Benchmarking attribute selection techniques for
  discrete class data mining,'' \emph{IEEE transactions on knowledge and data
  engineering}, vol.~15, no.~6, pp. 1437--1447, 2003.

\bibitem{Tant16a}
\BIBentryALTinterwordspacing
C.~Tantithamthavorn, S.~McIntosh, A.~E. Hassan, and K.~Matsumoto, ``Automated
  parameter optimization of classification techniques for defect prediction
  models,'' in \emph{Proceedings of the 38th International Conference on
  Software Engineering}, ser. ICSE '16.\hskip 1em plus 0.5em minus 0.4em\relax
  New York, NY, USA: ACM, 2016, pp. 321--332. [Online]. Available:
  \url{http://doi.acm.org/10.1145/2884781.2884857}
\BIBentrySTDinterwordspacing

\bibitem{Osma17a}
\BIBentryALTinterwordspacing
H.~Osman, M.~Ghafari, and O.~Nierstrasz, ``Hyperparameter optimization to
  improve bug prediction accuracy,'' in \emph{1st International Workshop on
  Machine Learning Techniques for Software Quality Evaluation (MaLTeSQuE
  2017)}, Feb. 2017, pp. 33--38. [Online]. Available:
  \url{http://scg.unibe.ch/archive/papers/Osma17a.pdf}
\BIBentrySTDinterwordspacing

\bibitem{Brei01a}
L.~Breiman, ``Random forests,'' \emph{Machine learning}, vol.~45, no.~1, pp.
  5--32, 2001.

\bibitem{Hsu03a}
C.-W. Hsu, C.-C. Chang, C.-J. Lin \emph{et~al.}, ``A practical guide to support
  vector classification,'' Department of Computer Science, National Taiwan
  University, Tech. Rep., 2003.

\bibitem{Guo04a}
L.~Guo, Y.~Ma, B.~Cukic, and H.~Singh, ``Robust prediction of fault-proneness
  by random forests,'' in \emph{Software Reliability Engineering, 2004. ISSRE
  2004. 15th International Symposium on}.\hskip 1em plus 0.5em minus
  0.4em\relax IEEE, 2004, pp. 417--428.

\bibitem{Elis08a}
K.~O. Elish and M.~O. Elish, ``Predicting defect-prone software modules using
  support vector machines,'' \emph{Journal of Systems and Software}, vol.~81,
  no.~5, pp. 649--660, 2008.

\bibitem{Flei71a}
J.~L. Fleiss, ``Measuring nominal scale agreement among many raters.''
  \emph{Psychological bulletin}, vol.~76, no.~5, p. 378, 1971.

\bibitem{Kaur08a}
A.~Kaur and R.~Malhotra, ``Application of random forest in predicting
  fault-prone classes,'' in \emph{2008 International Conference on Advanced
  Computer Theory and Engineering}.\hskip 1em plus 0.5em minus 0.4em\relax
  IEEE, 2008, pp. 37--43.

\bibitem{DeCa10a}
A.~B. De~Carvalho, A.~Pozo, and S.~R. Vergilio, ``A symbolic fault-prediction
  model based on multiobjective particle swarm optimization,'' \emph{Journal of
  Systems and Software}, vol.~83, no.~5, pp. 868--882, 2010.

\bibitem{Malh11a}
R.~Malhotra and Y.~Singh, ``On the applicability of machine learning techniques
  for object oriented software fault prediction,'' \emph{Software Engineering:
  An International Journal}, vol.~1, no.~1, pp. 24--37, 2011.

\bibitem{Zimm09a}
\BIBentryALTinterwordspacing
T.~Zimmermann, N.~Nagappan, H.~Gall, E.~Giger, and B.~Murphy, ``Cross-project
  defect prediction: A large scale experiment on data vs. domain vs. process,''
  in \emph{Proceedings of the the 7th Joint Meeting of the European Software
  Engineering Conference and the ACM SIGSOFT Symposium on The Foundations of
  Software Engineering}, ser. ESEC/FSE '09.\hskip 1em plus 0.5em minus
  0.4em\relax New York, NY, USA: ACM, 2009, pp. 91--100. [Online]. Available:
  \url{http://doi.acm.org/10.1145/1595696.1595713}
\BIBentrySTDinterwordspacing

\bibitem{Twal11a}
B.~Twala, ``Software faults prediction using multiple classifiers,'' in
  \emph{Computer Research and Development (ICCRD), 2011 3rd International
  Conference on}, vol.~4.\hskip 1em plus 0.5em minus 0.4em\relax IEEE, 2011,
  pp. 504--510.

\bibitem{Okut14a}
A.~Okutan and O.~T. Y{\i}ld{\i}z, ``Software defect prediction using bayesian
  networks,'' \emph{Empirical Software Engineering}, vol.~19, no.~1, pp.
  154--181, 2014.

\bibitem{Khos06a}
T.~M. Khoshgoftaar, N.~Seliya, and N.~Sundaresh, ``An empirical study of
  predicting software faults with case-based reasoning,'' \emph{Software
  Quality Journal}, vol.~14, no.~2, pp. 85--111, 2006.

\bibitem{Naga06a}
\BIBentryALTinterwordspacing
N.~Nagappan, T.~Ball, and A.~Zeller, ``Mining metrics to predict component
  failures,'' in \emph{Proceedings of the 28th international conference on
  Software engineering}, ser. ICSE '06.\hskip 1em plus 0.5em minus 0.4em\relax
  New York, NY, USA: ACM, 2006, pp. 452--461. [Online]. Available:
  \url{http://doi.acm.org/10.1145/1134285.1134349}
\BIBentrySTDinterwordspacing

\bibitem{Kanm07a}
S.~Kanmani, V.~R. Uthariaraj, V.~Sankaranarayanan, and P.~Thambidurai,
  ``Object-oriented software fault prediction using neural networks,''
  \emph{Information and software technology}, vol.~49, no.~5, pp. 483--492,
  2007.

\bibitem{Sing10a}
Y.~Singh, A.~Kaur, and R.~Malhotra, ``Prediction of fault-prone software
  modules using statistical and machine learning methods,'' \emph{International
  Journal of Computer Applications}, vol.~1, no.~22, pp. 8--15, 2010.

\bibitem{Menz07a}
T.~Menzies, J.~Greenwald, and A.~Frank, ``Data mining static code attributes to
  learn defect predictors,'' \emph{Software Engineering, IEEE Transactions on},
  vol.~33, no.~1, pp. 2--13, Jan. 2007.

\bibitem{Turh07a}
B.~Turhan and A.~Bener, ``A multivariate analysis of static code attributes for
  defect prediction,'' in \emph{Seventh International Conference on Quality
  Software (QSIC 2007)}.\hskip 1em plus 0.5em minus 0.4em\relax IEEE, 2007, pp.
  231--237.

\bibitem{Sing09a}
P.~Singh and S.~Verma, ``An investigation of the effect of discretization on
  defect prediction using static measures,'' in \emph{Advances in Computing,
  Control, \& Telecommunication Technologies, 2009. ACT'09. International
  Conference on}.\hskip 1em plus 0.5em minus 0.4em\relax IEEE, 2009, pp.
  837--839.

\bibitem{Misi11a}
A.~T. M{\i}s{\i}rl{\i}, A.~B. Bener, and B.~Turhan, ``An industrial case study
  of classifier ensembles for locating software defects,'' \emph{Software
  Quality Journal}, vol.~19, no.~3, pp. 515--536, 2011.

\bibitem{Cahi13a}
J.~Cahill, J.~M. Hogan, and R.~Thomas, ``Predicting fault-prone software
  modules with rank sum classification,'' in \emph{2013 22nd Australian
  Software Engineering Conference}.\hskip 1em plus 0.5em minus 0.4em\relax
  IEEE, 2013, pp. 211--219.

\bibitem{Gond08a}
I.~Gondra, ``Applying machine learning to software fault-proneness
  prediction,'' \emph{Journal of Systems and Software}, vol.~81, no.~2, pp.
  186--195, 2008.

\bibitem{Song11a}
Q.~Song, Z.~Jia, M.~Shepperd, S.~Ying, and J.~Liu, ``A general software
  defect-proneness prediction framework,'' \emph{IEEE Transactions on Software
  Engineering}, vol.~37, no.~3, pp. 356--370, 2011.

\bibitem{Turh09a}
B.~Turhan and A.~Bener, ``Analysis of naive bayes' assumptions on software
  fault data: An empirical study,'' \emph{Data \& Knowledge Engineering},
  vol.~68, no.~2, pp. 278--290, 2009.

\bibitem{Ostr05a}
T.~Ostrand, E.~Weyuker, and R.~Bell, ``Predicting the location and number of
  faults in large software systems,'' \emph{Software Engineering, IEEE
  Transactions on}, vol.~31, no.~4, pp. 340--355, Apr. 2005.

\bibitem{Jane06a}
A.~Janes, M.~Scotto, W.~Pedrycz, B.~Russo, M.~Stefanovic, and G.~Succi,
  ``Identification of defect-prone classes in telecommunication software
  systems using design metrics,'' \emph{Information sciences}, vol. 176,
  no.~24, pp. 3711--3734, 2006.

\bibitem{Rath16a}
S.~S. Rathore and S.~Kumar, ``An empirical study of some software fault
  prediction techniques for the number of faults prediction,'' \emph{Soft
  Computing}, pp. 1--18, 2016.

\bibitem{Grav00a}
T.~L. Graves, A.~F. Karr, J.~S. Marron, and H.~Siy, ``Predicting fault
  incidence using software change history,'' \emph{IEEE Transactions on
  Software Engineering}, vol.~26, no.~2, 2000.

\bibitem{Gao07a}
K.~Gao and T.~M. Khoshgoftaar, ``A comprehensive empirical study of count
  models for software fault prediction,'' \emph{IEEE Transactions on
  Reliability}, vol.~56, no.~2, pp. 223--236, 2007.

\end{thebibliography}
\end{document}